\documentclass[final,3p,times,square,sort&compress]{astyle}
\usepackage{txfonts}

\usepackage{amssymb}
\usepackage{CJKutf8} 
\usepackage{amsmath} 

\usepackage[dvipdfm,
            pdfstartview=FitH,
            CJKbookmarks=true,
            bookmarksnumbered=false,
            bookmarksopen=true,
            colorlinks=true, 
            citecolor=magenta,
            linkcolor=blue,
            linktocpage=true,
            ]{hyperref}     

\AtBeginDvi{}
 
\newtheorem{thm}{Theorem}[section]
\newproof{proof}{Proof}

\begin{document}
\begin{frontmatter}

\title{The Convolution Series Solution  of  Divergent  Navier-Stokes Equations
}
\author{ Yan Yimin
\footnote{Email:yanyimin@foxmail.com}
}
\address{}
\begin{abstract}
This paper gives out the solution of  divergent  Navier-Stokes equations, and shows that in this case,
under a physical acceptable condition, the solution would be smooth .
\end{abstract}
\end{frontmatter}



\section{Introduction}
Any physical system, the behavior of local state is actually a result of global interaction, which is probably  to  be interpreted as the  convolution of two objects
     \begin{equation*}
            f(\textbf{x})\ast g(\textbf{x})\doteq  \int_{\mathbb{R}^n}   f(\textbf{x-y })\cdot g(\textbf{y })d \textbf{y }
     \end{equation*}
 for instance:
\begin{enumerate}
  \item {Brownian motion of free electric charges:if $q(\textbf{x})$ is initial charge distribution  of the system ,  $ K(\textbf{x},t) = \frac{1}{(4 \pi t)^\frac{n}{2}}e^{-\frac{\textbf{x}^2 }{4 t}}$ is the heat conduction kernel, which means the probability of a charge move to a distance $\textbf{x}$ over a period of time $t$,
      then $  K(\textbf{x},t)\ast q(\textbf{x})  $ means the charge distribution after time $t$.
      Because  "probability" always makes us uneasy in understanding physical phenomenon, we give out another example.
  }
  \item {A electric system, if $q(\textbf{x})$ is the  charge distribution ,   $E(\textbf{x})$ is the  electric field, and the interaction between the field and the charge  is merely depended on the vector distant   $\textbf{x}$,
      then $  E(\textbf{x} )\ast q(\textbf{x})  $ means the total external force act on   each position.
  }
\end{enumerate}

It is not doubt that  convolution is of crucial important  in describing physical phenomenons.
To illustrate such idea, we provide a   convolution expression for Divergent Navier-Stokes Equations.

\section{Main Result}
\begin{thm}
For the   Navier-Stokes Equations with divergent  condition :
 \begin{equation} \label{NSE}
      \left\{
       \begin{aligned}
             &\partial_t \textbf{u}+(\textbf{u}\cdot \nabla)\textbf{u}=\triangle \textbf{u}+\nabla (f- p)\\
             &\textbf{u}(\textbf{x},t)|_{t=0}=\textbf{u}_0 (\textbf{x} )=\nabla \varphi(\textbf{x} )
       \end{aligned}
       \right.
  \end{equation}
  where $\nabla  f(\textbf{x},t)$ is the external force, $p(\textbf{x},t)$ is the pressure, and $ \varphi(\textbf{x} )$ is the initial potential  .\\
   If \begin{enumerate}
        \item { $ f(\textbf{x},t)- p(\textbf{x},t)\in L^\infty( \mathbb{R}^n \times [0,T])$  ; }
        \item {
              $\textbf{u}_0\in L^\infty( \mathbb{R}^n)$  , such as $\|\textbf{u}_0\|_\infty\leq$ light speed $c$, and integrable in any bounded closure of  $ \mathbb{R}^n$;
              }
         \item {
              $\exists\  \textbf{x}_0 \in \mathbb{R}^n$, s.t. $\varphi(\textbf{x}_0)=a$  is bounded.
          }
      \end{enumerate}
     then
\begin{enumerate}
  \item {
  the solution of such equations is
  \begin{equation}
  \textbf{u}( \textbf{x},t) =-2 \frac{\nabla G( \textbf{x},t)}{G( \textbf{x},t)}
  \end{equation}
  where
    \begin{equation}
      \left\{
       \begin{aligned}
              G( \textbf{x},t) & = K(\textbf{x},t)\ast e^{-\frac{1}{2} \varphi(\textbf{x}) }
              +  \int_0^t  K(\textbf{x},t-s)\ast  \Big[F(\textbf{x},s)[K(\textbf{x},s)\ast e^{-\frac{1}{2} \varphi(\textbf{x}) }]\Big] ds\\
              &+
                \int_0^t  K(\textbf{x},t-s)\ast \biggl[F(\textbf{x},s)
                \int_0^s  K(\textbf{x},s-\tau)\ast  \Big(F(\textbf{x},\tau)  [K(\textbf{x},\tau)\ast e^{-\frac{1}{2} \varphi(\textbf{x}) }] \Big) d\tau     \biggl] ds\\
                 &+\int_0^t  K(\textbf{x},t-s)\ast \biggl[F(\textbf{x},s)
                \int_0^s  K(\textbf{x},s-\tau)\ast  \biggl( F(\textbf{x},\tau)
                \int_0^\tau  K(\textbf{x},\tau-\theta)\ast  \Big[F(\textbf{x},\theta)[K(\textbf{x},\theta)\ast e^{-\frac{1}{2} \varphi(\textbf{x}) }]\Big] d\theta
                \biggl) d\tau    \biggl] ds\\
                &+\cdots \cdots\\
             K(\textbf{x},t)&= \frac{1}{(4 \pi t)^\frac{n}{2}}e^{-\frac{\textbf{x}^2 }{4 t}},\ \ \
             F(\textbf{x},t)=\frac{1}{2}   \big [  p(\textbf{x} ,t)-f(\textbf{x} ,t) \big ]
       \end{aligned}
       \right.
   \end{equation}

  }

  \item {If $\textbf{u}_0( \textbf{x} ), \textbf{F}( \textbf{x},t) \in C^{\infty}(\mathbb{R}^n\times [0,T])\bigcap L^{\infty}(\mathbb{R}^n\times [0,T])$, then 
  $    \textbf{u}( \textbf{x},t)  \in  C^{\infty}(\mathbb{R}^n\times [0,T])\bigcap L^{\infty}(\mathbb{R}^n\times [0,T])$.
  }

\end{enumerate}

\end{thm}

\begin{proof}

\subsection{Deduce the Expression of \textbf{u}(\textbf{x},t) }

It goes without saying that , if excluding  the nonlinear terms, Eq.(\ref{NSE})  turns out to be a heat conduction equation. So, take an undefined Kernel  $G(\textbf{x},t)$  into   consideration :
  \begin{equation} \label{test}
    G \partial_t \textbf{u} + G (\textbf{u}\cdot \nabla)\textbf{u}= G \triangle \textbf{u}+G\nabla(f- p)
  \end{equation}

  \begin{enumerate}
    \item {  Seeing that
    \begin{equation*}
            \partial_{xx} (G\ \textbf{u})= \partial_{xx}  G\cdot\textbf{u}+G\partial_{xx}   \textbf{u}   +2\partial_x G\cdot \partial_x\textbf{u}
    \end{equation*}
    so,
    \begin{equation*}
         \begin{aligned}
           \triangle(G\ \textbf{u})=&  \triangle G\cdot\textbf{u}+ G \triangle \textbf{u}
           +2\Big[\sum_{k=1}^n \partial_{x_k}  G\cdot \partial_{x_k}\textbf{u}  \Big ]\\
       =& \triangle G\cdot\textbf{u}+ G \triangle \textbf{u}+2[(\partial_{x_1}  G , \cdots,\partial_{x_n}  G )\cdot \nabla   ]\textbf{u} \\
       =& \triangle G\cdot\textbf{u}+ G \triangle \textbf{u}+2[ \nabla G\cdot \nabla   ]\textbf{u}
         \end{aligned}
    \end{equation*}

     therefore, Eq.(\ref{test}) is  equivalent to
     \begin{equation}\label{test.h}
         \begin{aligned}
              & \partial_t (G \ \textbf{u} )-\partial_t G\cdot \textbf{u}+(G \textbf{u}\cdot \nabla)\textbf{u}\\
       =  & \triangle(G\ \textbf{u})- \triangle G\cdot\textbf{u}  - 2[ \nabla G \cdot \nabla   ]\textbf{u} +G\nabla (f- p)
         \end{aligned}
    \end{equation}
    }
    \item {Now, assume that $G , \textbf{u}  $ satisfy that
    \begin{equation} \label{relation}
      \left\{
       \begin{aligned}
             &\partial_t G=\triangle G+Q(\textbf{x} ,t)\\
             &G\textbf{u} =-2\nabla G
       \end{aligned}
       \right.
   \end{equation}
    Then the answer occur to us that Eq.(\ref{test.h}) leaves out to be
    \begin{equation}
         -2\nabla Q -Q \textbf{u}=G \nabla (f-p)
    \end{equation}
    According to Eq.(\ref{relation}),
    \begin{equation}
        \textbf{u}=-2 \frac{\nabla G}{G}
    \end{equation}
    so,
    \begin{equation}
       \begin{aligned}
              \nabla (f-p)=-2\frac{\nabla Q}{G} - \frac{Q\textbf{u}}{G}
              =-2\frac{\nabla Q}{G} +2 \frac{Q}{G}   \frac{\nabla G}{G}
              =-2\nabla \frac{Q}{G}
       \end{aligned}
   \end{equation}
   Hence, it is clearly reasonable and compatible, if we set
   \begin{equation}\label{forcecontol}
      f-p=-2\frac{Q}{G}+C
   \end{equation}
   \emph{where $C=C(t)$. }
   As a result, G is the solution of
   \begin{equation}\label{extendHeatEqG}
      \partial_t G=\triangle G+\frac{1}{2}   \big [C+ p(\textbf{x} ,t)-f(\textbf{x} ,t) \big ]  G
   \end{equation}
   It has not escaped our attention that $C$ is not a essential  variable, since if we let $G=e^{\frac{1}{2} \int_0^t Cd\tau}\cdot W(\textbf{x} ,t)$, then
    \begin{equation*}
       \left\{
       \begin{aligned}
             & \textbf{u}( \textbf{x},t) =-2 \frac{\nabla G( \textbf{x},t)}{G( \textbf{x},t)}= -2 \frac{\nabla W( \textbf{x},t)}{W( \textbf{x},t)} \\
             &\partial_t W=\triangle W+\frac{1}{2}   \big [  p(\textbf{x} ,t)-f(\textbf{x} ,t) \big ]  W
       \end{aligned}
       \right.
	\end{equation*}
   So, let $C=0$.

    }
    \item {
     And let
     \begin{equation}
         F(\textbf{x},t)=\frac{1}{2}   \big [  p(\textbf{x} ,t)-f(\textbf{x} ,t) \big ]
     \end{equation}
    and consider \textbf{the controlled  heat conduction equation }

    \begin{equation}
      \left\{
       \begin{aligned}
            &\partial_t G( \textbf{x},t)=\triangle G( \textbf{x},t)+F( \textbf{x},t)\ G( \textbf{x},t)\\
            &G( \textbf{x},0) = G_0( \textbf{x} )
       \end{aligned}
       \right.
   \end{equation}
    which has the solution

    \begin{equation}\label{Control.Heat.Eq.1}
      \left\{
       \begin{aligned}
              G( \textbf{x},t) & = K(\textbf{x},t)\ast G_0(\textbf{x})
              +  \int_0^t  K(\textbf{x},t-s)\ast   \left(  F(\textbf{x},s)[ K(\textbf{x},s)\ast G_0(\textbf{x})]  \right)  ds\\
              &+
                \int_0^t  K(\textbf{x},t-s)\ast \biggl[F(\textbf{x},s)
                \int_0^s  K(\textbf{x},s-\tau)\ast
                \left(   F(\textbf{x},\tau)[  K(\textbf{x},\tau)\ast G_0(\textbf{x})]  \right)
                d\tau     \biggl] ds\\
             &+\int_0^t  K(\textbf{x},t-s)\ast \biggl[F(\textbf{x},s)
                \int_0^s  K(\textbf{x},s-\tau)\ast  \biggl( F(\textbf{x},\tau)
                \int_0^\tau  K(\textbf{x},\tau-\theta)\ast
                \left(  F(\textbf{x},\theta)[K(\textbf{x},\theta)\ast G_0(\textbf{x})]   \right)
                d\theta
                \biggl) d\tau    \biggl] ds\\
                &+\cdots \cdots\\
             K(\textbf{x},t)&= \frac{1}{(4 \pi t)^\frac{n}{2}}e^{-\frac{\textbf{x}^2 }{4 t}},\ \ \
            G( \textbf{x},0)= G_0(\textbf{x})
       \end{aligned}
       \right.
   \end{equation}

    Briefly  verification could be set up  by following steps:\\

    \begin{enumerate}
      \item {  It is clear that Eq.(\ref{Control.Heat.Eq.1}) is  equivalent  to
      \begin{equation}
         G( \textbf{x},t)=K(\textbf{x},t)\ast G_0(\textbf{x})+  \int_0^t  K(\textbf{x},t-s)\ast
              \biggl(  F(\textbf{x},s)\cdot G( \textbf{x},s) \biggl)  ds
      \end{equation}

   then
	\begin{equation*}
       \begin{aligned}
            \partial_t G( \textbf{x},t)
            &=\partial_t K(\textbf{x},t)\ast G_0(\textbf{x})+ \partial_t  \int_0^t  K(\textbf{x},t-s)\ast
             \biggl(  F(\textbf{x},s)\cdot G( \textbf{x},s) \biggl) ds\\
           & =\partial_t K(\textbf{x},t)\ast G_0(\textbf{x})
             +\int_0^t \partial_t K(\textbf{x},t-s)\ast
               \biggl(  F(\textbf{x},s)\cdot G( \textbf{x},s)\biggl) ds
               +\lim_{s \rightarrow t} K(\textbf{x},t-s )\ast
              \biggl(  F(\textbf{x},s)\cdot G( \textbf{x},s) \biggl)\\
           & =\partial_t K(\textbf{x},t)\ast G_0(\textbf{x})
            +\int_0^t \Delta K(\textbf{x},t-s)\ast
               \biggl(  F(\textbf{x},s)\cdot G( \textbf{x},s)\biggl) ds
               + F(\textbf{x},t)\cdot G( \textbf{x},t)\\
            &=  \Delta  K(\textbf{x},t)\ast G_0(\textbf{x})
            +\Delta \int_0^t K(\textbf{x},t-s)\ast
               \biggl(  F(\textbf{x},s)\cdot G( \textbf{x},s) \biggl) ds
               + F(\textbf{x},t)\cdot G( \textbf{x},t)  \\
            &=   \Delta G( \textbf{x},t) + F(\textbf{x},t)\ G( \textbf{x},t)
       \end{aligned}
   \end{equation*}

      }
      \item {Beside, let's show that $G( \textbf{x},t)$ is  absolute convergence for all fixed point $( \textbf{x},t)\in \mathbb{R}^n \times [0,T]$  .\\
  if   $F( \textbf{x},t)$ is bounded, i.e.
  \begin{center}
   $\exists\ M>0$, \quad s.t.\quad
  $  \|F( \textbf{x},t)\|_{L^\infty( \mathbb{R}^n \times [0,T])}\leq M $.
 \end{center}

  then,
    \begin{equation*}
       \begin{aligned}
             \Big | G( \textbf{x},t) \Big |
             &
             \leq \Big |K(\textbf{x},t)\ast G_0(\textbf{x})\Big |
              +  \Big |\int_0^t  K(\textbf{x},t-s)\ast   \left(  F(\textbf{x},s)[ K(\textbf{x},s)\ast G_0(\textbf{x})]  \right)  ds\Big |\\
              &+
                \Big |\int_0^t  K(\textbf{x},t-s)\ast \biggl[F(\textbf{x},s)
                \int_0^s  K(\textbf{x},s-\tau)\ast
                \left(   F(\textbf{x},\tau)[  K(\textbf{x},\tau)\ast G_0(\textbf{x})]  \right)
                d\tau     \biggl] ds\Big |\\
             &+\Big |\int_0^t  K(\textbf{x},t-s)\ast \biggl[F(\textbf{x},s)
                \int_0^s  K(\textbf{x},s-\tau)\ast  \biggl( F(\textbf{x},\tau)
                \int_0^\tau  K(\textbf{x},\tau-\theta)\ast
                \left(  F(\textbf{x},\theta)[K(\textbf{x},\theta)\ast G_0(\textbf{x})]   \right)
                d\theta
                \biggl) d\tau    \biggl] ds\Big |\\
                &+\cdots \cdots\\
           &\leq   K(\textbf{x},t)\ast \Big |G_0(\textbf{x})\Big |
              +   \int_0^t  K(\textbf{x},t-s)\ast   \left( M  K(\textbf{x},s)\ast \Big |G_0(\textbf{x})\Big |   \right)  ds  \\
              &+
                 \int_0^t  K(\textbf{x},t-s)\ast \biggl[M
                \int_0^s  K(\textbf{x},s-\tau)\ast
                \left(  M   K(\textbf{x},\tau)\ast\Big | G_0(\textbf{x})\Big |   \right)
                d\tau     \biggl] ds \\
             &+\int_0^t  K(\textbf{x},t-s)\ast \biggl[M
                \int_0^s  K(\textbf{x},s-\tau)\ast  \biggl( M
                \int_0^\tau  K(\textbf{x},\tau-\theta)\ast
                \left(  M K(\textbf{x},\theta)\ast\Big | G_0(\textbf{x})\Big |    \right)
                d\theta
                \biggl) d\tau    \biggl] ds  +\cdots \cdots\\
         &  \stackrel{def}{=}
         V( \textbf{x},t)
       \end{aligned}
   \end{equation*}

   Clearly, $V( \textbf{x},t)$ is the solution of
    \begin{equation}
      \left\{
       \begin{aligned}
            &\partial_t V( \textbf{x},t)=\triangle V( \textbf{x},t)+M\ V( \textbf{x},t)\\
            &V( \textbf{x},0) = |G_0( \textbf{x} )|
       \end{aligned}
       \right.
   \end{equation}

   so, let $V( \textbf{x},t)=e^{Mt}\cdot W( \textbf{x},t)$, we get
    \begin{equation*}
      \left\{
       \begin{aligned}
            &\partial_t W( \textbf{x},t)=\triangle W( \textbf{x},t) \\
            &W( \textbf{x},0) = |G_0( \textbf{x} )|
       \end{aligned}
       \right.
   \end{equation*}
   i.e.
   \begin{equation*}
    W( \textbf{x},t)=K(\textbf{x},t)\ast \Big |G_0(\textbf{x})\Big |
   \end{equation*}

   Therefore, we get  a fundamental conclusion
   \begin{equation} \label{GUpperBonuded}
    \Big | G( \textbf{x},t) \Big |
    \leq e^{Mt}\cdot K(\textbf{x},t)\ast \Big |G_0(\textbf{x})\Big |
   \end{equation}

    So,  if $G_0(\textbf{x})$ are  bounded, such as $|G_0(\textbf{x})|\leq M$, it follows
      \begin{equation*}
        \Big | G( \textbf{x},t) \Big |
        \leq e^{Mt}\cdot K(\textbf{x},t)\ast M=M e^{Mt}
      \end{equation*}
     But  such restriction  is too narrow.   In the next step, we will  consider a general case , so that  $G( \textbf{x},t)$ is still absolute convergence at any   fixed point $( \textbf{x},t)\in \mathbb{R}^n \times [0,T]$ .
     }
    \end{enumerate}

    }
   \item {Firstly,
    \begin{equation}
            G_0(\textbf{x})= e^{-\frac{1}{2}   \varphi(\textbf{x}) }
     \end{equation}
     because according to the Assumption(\ref{relation}):
    \begin{equation}
        \textbf{u}=-2 \frac{\nabla G}{G}=-2\nabla ln \ G
    \end{equation}
    it follows
     \begin{equation*}
       -2 \triangle ln \ G_0= div\ \textbf{u}_0=  div \ \nabla \varphi(\textbf{x}) =\triangle \varphi(\textbf{x})
    \end{equation*}

    or
      \begin{equation*}
         -2  ln \ G_0=\varphi(\textbf{x})+ g(\textbf{x}),\qquad  \triangle  g(\textbf{x})=0
      \end{equation*}
   So, if we add an additional restriction that:
   \begin{equation*}
     \lim_{|\textbf{x}|\rightarrow \infty} G_0(\textbf{x}) \quad \hbox{ and  }  \quad
      \lim_{|\textbf{x}|\rightarrow \infty} \partial_{x_i} G_0(\textbf{x})  \hbox{ are possible  bounded }
   \end{equation*}
    we get
    \begin{equation*}
        ln \ G_0     =  -\frac{1}{2}   \varphi(\textbf{x})\quad \hbox{ or }
        \quad  G_0=e^{-\frac{1}{2}   \varphi(\textbf{x})}
           , \qquad g(\textbf{x})=0
    \end{equation*}\\


    Secondly, if the initial condition meets  the \textbf{ Restricted Conditions} ,
    \begin{itemize}\label{BasicRestriction}
      \item {\texttt{Bounded Restriction}: \quad
      $\textbf{u}_0$  is bounded, such as $|\textbf{u}_0|\leq$ light speed $c$, and integrable in any bounded closure of $\mathbb{R}^n$
      }
      \item {\texttt{Non-trivial Restriction:} \quad
      $\exists\  \textbf{x}_0 \in \mathbb{R}^n$, s.t. $\varphi(\textbf{x}_0)=a$  is bounded;  \\
         otherwise, $\forall \ \textbf{x}  \in \mathbb{R}^n$,  $\varphi(\textbf{x} )=\infty \hbox{ or } -\infty $ or non-existing, which is a trivial case we won't  consider.\\
          Without lose of generality, assume $\textbf{x}_0 =0, $ i.e. $ \varphi( 0)=a$  is bounded.\\
      }
    \end{itemize}

        Then    it is clear that, in the non-rotational fluid,  the integral along any  curve  $\quad L: \ \textbf{x}_0 \rightarrow \textbf{x}   $
          \begin{equation*}
             \varphi(\textbf{x})
             =\int_{L}  \textbf{u}_0\ d \textbf{x} +a
          \end{equation*}
          is independent of the path.

          So
           \begin{equation}\label{Estimation.G.upper.begin}
               \begin{aligned}
        	     e^{-\frac{1}{2}   \varphi(\textbf{x}) }
                 =e^{-\frac{1}{2} \Big(\int_{L}  \textbf{u}_0 d \textbf{x} +a \Big)   }
                \leq
                e^{ \frac{1}{2} \int_{0}^{r}  c\ d r   -\frac{a}{2}  }
                 =e^{ \frac{c  r -a}{2} }
                \qquad \Big( \ r =|\textbf{x}-\textbf{x}_0|    \ \Big)
               \end{aligned}
           \end{equation}

           Consider the  spherical coordinates $(r,\theta,\phi)$:
                \begin{equation*}
                   \left\{
                   \begin{aligned}
                         & x=r \cos \theta  \sin \phi\\
                         & y=r \sin \theta  \sin \phi\\
                         & z=r \cos   \phi
                         \qquad \qquad \Big(\
                         r\in [0,\infty),\ \theta\in[0,2\pi],\ \phi\in[0,\pi]\
                         \Big)
                   \end{aligned}
                   \right.
            	\end{equation*}
           then

           \begin{equation*}
               \begin{aligned}
        	      K(\textbf{x},t )\ast e^{-\frac{1}{2}   \varphi(\textbf{x}) }
                =\int_{\mathbb{R}^3} \frac{1}{(4 \pi t)^\frac{3}{2}}e^{-\frac{(\textbf{x}-\textbf{y} )^2 }{4 t}}
                e^{-\frac{1}{2}   \varphi(\textbf{y}) }  d\textbf{y}
                =\int_{0}^{2\pi}d\theta
                \int_{0}^{ \pi}d \phi
                \int_{0}^{\infty }d  \rho \bigg(\rho^2 \sin \phi
                  \frac{1}{(4 \pi t)^\frac{3}{2}}e^{-\frac{(r^2+\rho^2-2b r\rho    )  }{4 t}}
                e^{-\frac{1}{2}   \varphi(\textbf{y}) }
                \bigg)
               \end{aligned}
           \end{equation*}
           \emph{where}
           \begin{equation*}
               \begin{aligned}
        	      b&=(\cos \theta  \sin \phi)\cdot (\cos \theta_1  \sin \phi_1)+
                     (\sin \theta  \sin \phi)\cdot (\sin \theta_1  \sin \phi_1)+
                     ( \cos   \phi  )\cdot ( \cos   \phi_1) \\
                    & \leq
                     \Big[
                     (\cos \theta  \sin \phi)^2+(\sin \theta  \sin \phi)^2+( \cos   \phi  )^2
                      \Big]^\frac{1}{2}
                      \cdot
                      \Big[
                      (\cos \theta_1  \sin \phi_1)^2+(\sin \theta_1  \sin \phi_1)^2+( \cos   \phi_1)^2
                      \Big]^\frac{1}{2}
                      =1
               \end{aligned}
           \end{equation*}

           Therefore
           \begin{equation}\label{Upper.Worst.InitialCondition}
               \begin{aligned}
        	      K(\textbf{x},t )\ast e^{-\frac{1}{2}   \varphi(\textbf{x}) }
                &\leq
                 \int_{0}^{2\pi}d\theta
                \int_{0}^{ \pi}d \phi
                \int_{0}^{\infty }d  \rho \bigg(\rho^2 \sin \phi
                  \frac{1}{(4 \pi t)^\frac{3}{2}}e^{-\frac{(r^2+\rho^2-2  r\rho    )  }{4 t}}
                e^{ \frac{c \rho  -a}{2} }
                \bigg)
                =4\pi
                \int_{0}^{\infty }  \bigg(\rho^2
                  \frac{1}{(4 \pi t)^\frac{3}{2}}e^{-\frac{(r^2+\rho^2-2  r\rho    )  }{4 t}}
                e^{ \frac{c \rho  -a}{2} }
                \bigg)d\rho\\
                &
                =
                e^{-\frac{ (r^2-B^2 ) }{4 t}- \frac{a}{2}}
                4\pi\frac{1}{(4 \pi t)^\frac{3}{2}}
                \int_{0}^{\infty }   \rho^2
                  e^{-\frac{( \rho -B )^2  }{4 t}}
                 d\rho
                 \qquad \qquad  \big(\  B=r+c t     \  \big)\\
                &=
                e^{ \frac{ t c^2 }{4}+ \frac{r c-a}{2}}
                 \frac{4}{\sqrt{\pi}  }
                \int_{-\eta}^{\infty }   (\xi+\eta)^2
                  e^{- \xi^2 }
                 d\xi
                 \qquad \qquad  \big(\  \eta=   \frac{B}{ \sqrt{4t}}= \frac{r+c t}{ \sqrt{4t}} ,\quad
                 \xi=\frac{ \rho }{ \sqrt{4t}} -\eta
                    \  \big)\\
              & \leq
              e^{ \frac{ t c^2 }{4}+ \frac{r c-a}{2}}
                 \frac{4}{\sqrt{\pi}  }
                \int_{-\infty}^{\infty }   (\xi+\eta)^2
                  e^{- \xi^2 }
                 d\xi
              =e^{ \frac{ t c^2 }{4}+ \frac{r c-a}{2}}
                 \frac{4}{\sqrt{\pi}  }
                 \cdot
                 \sqrt{\pi} \Big(\eta^2+\frac{1}{2}  \Big)\\
                &=
                e^{ \frac{ t c^2 }{4}+ \frac{r c-a}{2}}
                 \cdot
                  4 \Big(\eta^2+\frac{1}{2}  \Big)
                =  e^{ \frac{ t c^2 }{4}+ \frac{r c-a}{2}}
                 \cdot
                    \Big[ \frac{(r+c t)^2}{t}+2 \Big]
               \end{aligned}
           \end{equation}
      which is clearly bounded   at any    fixed point $( \textbf{x},t)\in \mathbb{R}^3 \times [0,T] $, so it is
      \begin{equation}
        | G( \textbf{x},t) |
                \leq e^{Mt}\cdot K(\textbf{x},t)\ast e^{-\frac{1}{2}   \varphi(\textbf{x}) }
      \end{equation} \\\\

   }
  \end{enumerate}


 As a result, we get the desired expression  
    \begin{equation}
      \left\{
       \begin{aligned}
       \textbf{u}( \textbf{x},t)&=-2 \frac{\nabla G( \textbf{x},t)}{G( \textbf{x},t)}\\
              G( \textbf{x},t) & = K(\textbf{x},t)\ast e^{-\frac{1}{2} \varphi(\textbf{x}) }
              +  \int_0^t  K(\textbf{x},t-s)\ast   \left(  F(\textbf{x},s)[ K(\textbf{x},s)\ast e^{-\frac{1}{2} \varphi(\textbf{x}) }]  \right)  ds\\
              &+
                \int_0^t  K(\textbf{x},t-s)\ast \biggl[F(\textbf{x},s)
                \int_0^s  K(\textbf{x},s-\tau)\ast
                \left(   F(\textbf{x},\tau)[  K(\textbf{x},\tau)\ast e^{-\frac{1}{2} \varphi(\textbf{x}) } ]  \right)
                d\tau     \biggl] ds\\
             &+\int_0^t  K(\textbf{x},t-s)\ast \biggl[F(\textbf{x},s)
                \int_0^s  K(\textbf{x},s-\tau)\ast  \biggl( F(\textbf{x},\tau)
                \int_0^\tau  K(\textbf{x},\tau-\theta)\ast
                \left(  F(\textbf{x},\theta)[K(\textbf{x},\theta)\ast e^{-\frac{1}{2} \varphi(\textbf{x}) }]   \right)
                d\theta
                \biggl) d\tau    \biggl] ds\\
                &+\cdots \cdots
       \end{aligned}
       \right.
   \end{equation}
   . \\\\

\subsection{\textbf{u  }(\textbf{x},t) in  $C^{\infty} (\mathbb{R}^n \times [0,T])  $  }
In order to obtain  $ \textbf{u}( \textbf{x},t) =-2 \frac{\nabla G( \textbf{x},t)}{G( \textbf{x},t)} \in  C^{\infty} (\mathbb{R}^n\times [0,T])$,
  all we need to prove are
 \begin{enumerate}
   \item { $\forall\ (\textbf{x},t)\in \mathbb{R}^n\times [0,T] ,  \exists \ \varepsilon_{x,t}>0 \ ,\quad s.t.\quad  G( \textbf{x},t)>\varepsilon_{xt}$.  }
   \item {
     $G( \textbf{x},t)\in    C^{\infty} ( \mathbb{R}^n \times [0,T]) $  }
 \end{enumerate}
so,

\subsubsection{Floor Estimation of G( \textbf{x},t) }\label{BoundEsSction}

For $G( \textbf{x},t)$, its bound could be estimated as follows:
\begin{enumerate}
  \item
  {If $F(\textbf{x} ,t)\geq 0 $,
   \begin{equation*}\label{Item.bound.1}
       \begin{aligned}
        G( \textbf{x},t)   = &K(\textbf{x},t)\ast e^{-\frac{1}{2} \varphi(\textbf{x}) }
              +  \int_0^t  K(\textbf{x},t-s)\ast   \left(  F(\textbf{x},s)[ K(\textbf{x},s)\ast e^{-\frac{1}{2} \varphi(\textbf{x}) }]  \right)  ds\\
              &+
                \int_0^t  K(\textbf{x},t-s)\ast \biggl[F(\textbf{x},s)
                \int_0^s  K(\textbf{x},s-\tau)\ast
                \left(   F(\textbf{x},\tau)[  K(\textbf{x},\tau)\ast e^{-\frac{1}{2} \varphi(\textbf{x}) } ]  \right)
                d\tau     \biggl] ds\\
             &+\int_0^t  K(\textbf{x},t-s)\ast \biggl[F(\textbf{x},s)
                \int_0^s  K(\textbf{x},s-\tau)\ast  \biggl( F(\textbf{x},\tau)
                \int_0^\tau  K(\textbf{x},\tau-\theta)\ast
                \left(  F(\textbf{x},\theta)[K(\textbf{x},\theta)\ast e^{-\frac{1}{2} \varphi(\textbf{x}) }]   \right)
                d\theta
                \biggl) d\tau    \biggl] ds\\
                &+\cdots \cdots\\
          \geq &
               K(\textbf{x},t)\ast e^{-\frac{1}{2} \varphi(\textbf{x}) }
       \end{aligned}
   \end{equation*}
  }
  \item {For the general case: $F(\textbf{x} ,t)$ is no always non-negative, but
  \begin{equation*}
      \inf_{\textbf{x}\in \mathbb{R}^n} \big \{F(\textbf{x} ,t) \big \}
     \qquad \hbox{ is  exsting    and integrable in } [0,T]
  \end{equation*}

      Considering the transformation
       \begin{equation*}
        G(\textbf{x} ,t)=e^{\int_0^{t} \inf_{\textbf{x}\in \mathbb{R}^n} \big \{F(\textbf{x} ,s) \big \} ds} \cdot W(\textbf{x} ,t)
       \end{equation*}
      so
   \begin{equation*}
      \left\{
       \begin{aligned}
             &\partial_t \ G=\triangle \ G+F(\textbf{x} ,t)\ G \\
             &G(\textbf{x} ,0) =K(\textbf{x},t)\ast e^{-\frac{1}{2} \varphi(\textbf{x}) }
       \end{aligned}
       \right.
   \end{equation*}
   turns out to be
    \begin{equation*}
      \left\{
       \begin{aligned}
             &\partial_t \ W=\triangle \ W+[F(\textbf{x} ,t)-
             \inf_{\textbf{x}\in \mathbb{R}^n} \big \{F(\textbf{x} ,t) \big \}]\ W
             \qquad \quad
             \Big(\ F(\textbf{x} ,t)- \inf_{\textbf{x}\in \mathbb{R}^n} \big \{F(\textbf{x} ,t) \big \}
             \geq 0 \ \Big)  \\
             &W(\textbf{x} ,0)=G(\textbf{x} ,0) =K(\textbf{x},t)\ast e^{-\frac{1}{2} \varphi(\textbf{x}) }
       \end{aligned}
       \right.
   \end{equation*}

  According to Case (\ref{Item.bound.1}), we get
  \begin{equation*}
     W(\textbf{x} ,t)\geq K(\textbf{x},t)\ast e^{-\frac{1}{2} \varphi(\textbf{x}) }
  \end{equation*}
  }
\end{enumerate}
Therefore,
\begin{equation}\label{Estimation.G.floor}
    G(\textbf{x} ,t)
   = e^{\int_0^{t} \inf_{\textbf{x}\in \mathbb{R}^n} \big \{F(\textbf{x} ,s) \big \} ds} \cdot W(\textbf{x} ,t)
    \geq e^{\int_0^{t}
     \inf_{\textbf{x}\in \mathbb{R}^n} \big \{F(\textbf{x} ,s) \big \}
     ds} \cdot \Big[ K(\textbf{x},t)\ast e^{-\frac{1}{2} \varphi(\textbf{x}) }\Big]
\end{equation}
\emph{Similarly,  we can obtain the upper estimation:}
\begin{equation*}
    G(\textbf{x} ,t)
   \leq
   e^{\int_0^{t}\Big ( \inf_{\textbf{x}\in \mathbb{R}^n} \big \{F(\textbf{x} ,s) \big \}+
   \sup_{\textbf{x}\in \mathbb{R}^n} \big \{F(\textbf{x} ,s) \big \}
    \Big )ds  }
    \cdot
     \Big[ K(\textbf{x},t)\ast e^{-\frac{1}{2} \varphi(\textbf{x}) }\Big]
  \leq
     e^{\int_0^{t}
  2 \sup_{\textbf{x}\in \mathbb{R}^n} \big \{F(\textbf{x} ,s) \big \}
    ds  }
    \cdot
     \Big[ K(\textbf{x},t)\ast e^{-\frac{1}{2} \varphi(\textbf{x}) }\Big]
\end{equation*}  \\

So, we can get a simple estimation  :
 \begin{enumerate}
   \item {according to  our Restriction \ref{BasicRestriction}, $\forall$ bounded closure   $\mho$, $\varphi(\textbf{x})   =\int_{L}  \textbf{u}_0\ d \textbf{x} +a$ is bounded in  $\mho$,  i.e. $\exists\ M_0>0$, s.t.\quad $e^{-\frac{1}{2} \varphi(\textbf{x}) }  \geq M_0  $  in $\mho$.
       Then,  $\forall\ (\textbf{x},t)\in \mathbb{R}^n\times [0,T]$,\quad $\exists\ \delta_{xt} >0$,  s.t.
       \begin{equation*}
         K(\textbf{x},t)\ast e^{-\frac{1}{2} \varphi(\textbf{x}) }
         =\int_{ \mathbb{R}^3}  K(\textbf{x}-\textbf{y},t)\cdot e^{-\frac{1}{2} \varphi(\textbf{y}) }d\textbf{y}
         \geq
         \int_{\mho}  K(\textbf{x}-\textbf{y},t)\cdot e^{-\frac{1}{2} \varphi(\textbf{y}) }d\textbf{y}
         \geq
         \int_{\mho}  K(\textbf{x}-\textbf{y},t)\cdot M_0   d\textbf{y}
         \stackrel{def}{=} \delta_{xt}
       \end{equation*}
   }
   \item {Since $ F(\textbf{x} ,t)  $  are bounded  , i.e.
         \begin{equation}
             \exists \ M_1>0,\quad s.t. \quad
              \inf_{\textbf{x}\in \mathbb{R}^n} \big \{F(\textbf{x} ,t) \big \}  \geq -M_1
         \end{equation}
        then, $\exists \  \varepsilon_{xt}>0$, s.t.
        \begin{equation}
            G(\textbf{x} ,t)
            \geq
            e^{\int_0^{t}-M_1 ds}\cdot
             K(\textbf{x},t)\ast e^{-\frac{1}{2} \varphi(\textbf{x}) }
            \geq
             e^{ -M_1 \ T} \cdot   \delta_{xt}
              \stackrel{def}{=}
                \varepsilon_{xt}
               \qquad
               \quad \Big( (\textbf{x} , t)\in  \mathbb{R}^n\times [0,T] \Big)
        \end{equation}
   }
 \end{enumerate}

In all , $\forall\ (\textbf{x} , t)\in  \mathbb{R}^n\times [0,T],\  \exists \ \varepsilon_{xt}>0 \ ,\quad s.t.\quad  G( \textbf{x},t)>\varepsilon_{xt}$.

\subsubsection{$G(\textbf{x},t)$ is Smooth with Respect to \textbf{x}}
Recalling that:
\begin{equation*}
    G( \textbf{x},t)=K(\textbf{x},t)\ast e^{-\frac{1}{2}   \varphi(\textbf{x}) }+  \int_0^t  K(\textbf{x},t-s)\ast [F(\textbf{x},s)G(\textbf{x},s) ] ds
\end{equation*}
Now one can   assert  that:
\begin{center}
 $G( \textbf{x},t)\in C ^{\infty}(\mathbb{R}^n \times [0,T] )$   , if $\textbf{u}_0\in C ^{\infty}(\mathbb{R}^n) \bigcap L^{\infty}(\mathbb{R}^n)$,  $\textbf{F}(\textbf{x},t)\in C ^{\infty}(\mathbb{R}^n\times [0,T]) \bigcap L^{\infty}(\mathbb{R}^n\times [0,T])$  .
\end{center}

Because $\forall\ \alpha  \in \mathbb{N}^n$ w.r.t. $\textbf{x}$:
\begin{enumerate}
  \item{  $\partial^\alpha G(\textbf{x},t)$ \textbf{bounded and  exists} . \\
 By Mathematical induction, we can assume  $\partial^ \beta  G(\textbf{x},t)$ is bounded ($\forall\ |\beta|<|\alpha|$), then
 \begin{equation*}
     \partial^\alpha (FG)\lesssim  F\cdot \partial^\alpha  G +1\lesssim 1+ |\partial^\alpha  G|
 \end{equation*}
So,  
   \begin{equation*}
       \begin{aligned}
	  |\partial^\alpha  G|
       &\lesssim
       1+  \int_0^t  K(\textbf{x},t-s)\ast |\partial^\alpha  G| ds
       \leq
       1+  \int_0^t  K(\textbf{x},t-s) \ast 1\ ds
       +\int_0^t  K(\textbf{x},t-s) \ast \int_0^s  K(\textbf{x}, s-\tau)\ast |\partial^\alpha  G| d\tau ds\\
       &\leq \cdots
       \leq
       1+  \int_0^t  K(\textbf{x},t-s) \ast 1\ ds
       +\int_0^t  K(\textbf{x},t-s) \ast \int_0^s  K(\textbf{x}, s-\tau)\ast 1\ d\tau ds\\
     &\qquad\qquad  +\int_0^t  K(\textbf{x},t-s) \ast \int_0^s  K(\textbf{x}, s-\tau)\ast \int_0^\tau  K(\textbf{x},\tau -\zeta) \ast 1\ d\zeta  d\tau ds+\cdots\\
      & =1+t+\frac{t^2}{2!}+\frac{t^3}{3!}+\cdots
       =e^t
       \end{aligned}
   \end{equation*} 

  Therefore,  $\partial^\alpha G(\textbf{x},t)$ bounded and  exists .
  }

  \item{$ \partial^\alpha G(\textbf{x},t)$ \textbf{is continuous}. 
   }
\end{enumerate}
Therefore, $G(\textbf{x},t )\in C^{\infty}(\mathbb{R}^n)\bigcap L^{\infty}(\mathbb{R}^n )$ . Further,  $G(\textbf{x},t )\in C^{\infty}(\mathbb{R}^n\times [0,T])\bigcap L^{\infty}(\mathbb{R}^n\times [0,T])$.

\bigskip\bigskip

Now, it is all clearly: for $\textbf{u}( \textbf{x},t) =-2 \frac{\nabla G( \textbf{x},t)}{G( \textbf{x},t)}$ ,
\begin{equation}
           \forall\ (\textbf{x} , t)\in  \mathbb{R}^n\times [0,T], \qquad
           \textbf{u}( \textbf{x},t) =-2 \frac{\nabla G( \textbf{x},t)}{G( \textbf{x},t)}
           \quad \in  C^{\infty}(\mathbb{R}^n\times [0,T])\bigcap L^{\infty}(\mathbb{R}^n\times [0,T])
 \end{equation}

\end{proof}

\section{Solution of One Dimension Parabolic Equation}
The  convolution series plays a crucial important role in solving PDEs. Here we give another example  as illustration.
Our following discussion bases on N.H. Ibragimov's work\cite{NHI} on   one dimension parabolic equations:
 \begin{equation} \label{1ProEq.eq}
   u_t+A(t,x)\cdot u_{xx}+a(t,x)\cdot u_x+c(t,x)\cdot u+f(t,x)=0\qquad
  (\ A(t,x)\leq 0\ )
 \end{equation}

\begin{verse}  \itshape         
 Let
 \begin{equation*}
     \tau=\phi(t), \quad
     y=\psi(t,x)
 \end{equation*}
 then it follows
     \begin{equation*}
       \left\{
       \begin{aligned}
             &u_t=\phi_t u_\tau+\psi_t u_y  \\
             &u_x=\psi_x u_y\\
             &u_{xx}=\psi_x^2 u_{yy}+\psi_{xx} u_y
       \end{aligned}
       \right.
	\end{equation*}
   so Eq.(\ref{1ProEq.eq}) becomes
   \begin{equation*}
     \phi_t u_\tau+A \psi_x^2 u_{yy}+(\psi_t+A \psi_{xx}+a\psi_{x})u_y+cu+f=0
   \end{equation*}
   Therefore, by choosing $\phi,\psi$, s.t.
   \begin{equation*}
     \phi_t=1,\qquad
     A\cdot \psi_x^2=-1
   \end{equation*}
   i.e.
   \begin{equation*}
     \tau=t,\qquad
     \psi_x=\sqrt{-\frac{1}{A}}
   \end{equation*}
   then we get
   \begin{equation} \label{1ProEq.eq2}
       u_t- u_{yy}+P\cdot u_y+cu+f=0 \qquad
       (P=\psi_t+A \psi_{xx}+a\psi_{x})
   \end{equation}\\

   Employ the transformation
    $$ v= e^{-\rho(t,y)}\cdot u$$
    it follows that
    \begin{equation*}
       u_t- u_{yy}+P\cdot u_y+cu+f
       =\big[
        v_t- v_{yy}+(P+2\rho_y)\cdot v_y
        +(\rho_{yy}-\rho_y^2-\rho_t-P\cdot\rho_y+  c)v
       \big] \cdot e^{-\rho(t,y)}+f
    \end{equation*}
    Clearly, if
    \begin{equation*}
      P+2\rho_y=0,\qquad or \quad  \rho=-\frac{1}{2}\int_0^y P(t,z)dz
    \end{equation*}
    Eq.(\ref{1ProEq.eq2}) turns out to be
    \begin{equation} \label{1ProEq.eq3}
          v_t- v_{yy}+Q\cdot v+g=0
    \end{equation}
    where
    \begin{equation*}
         Q=\rho_{yy}-\rho_y^2-\rho_t-P\cdot\rho_y+  c
         =-\frac{1}{2}P_y+\frac{1}{4}P^2+\frac{1}{2}\int_0^y P_t dy+c,
         \qquad
         g=f\cdot e^{-\frac{1}{2}\int_0^y P(t,z)dz }
    \end{equation*}

\end{verse}
Now, it is clearly that Eq.(\ref{1ProEq.eq3}) is equitant to
\begin{equation}
    v(x,t)=K\ast v_0- \int_0^t K(x,t-s)\ast\big[ Q(x,s)\cdot v(x,s)+g(x,s)\big]ds
\end{equation}
whose solution could be easily obtained by irritating.\\

Besides, it is   convenient to give the notion
  \begin{equation*}
   v(x,t) :=G+\mathcal{L}v(x,t)=
   \bigg[K\ast v_0- \int_0^t K(x,t-s)\ast g(x,s) ds\bigg]
    -\int_0^t K(x,t-s)\ast\big[ Q(x,s)\cdot v(x,s) \big]ds
  \end{equation*}
  so that we have the expression

  \begin{equation}
       \begin{aligned}
	       v&=G+\mathcal{L}v\\
            & =G+\mathcal{L}[G+\mathcal{L}v ] =G+\mathcal{L}G+\mathcal{L} \mathcal{L}v  \\
            &=G+\mathcal{L}G+\mathcal{L} \mathcal{L}G+\mathcal{L} \mathcal{L}\mathcal{L}v\\
            &=\cdots\cdots\\
            &= G+\mathcal{L}G+\mathcal{L} \mathcal{L}G+\mathcal{L} \mathcal{L}\mathcal{L}G+\cdots\cdots
       \end{aligned}
   \end{equation}

\bibliographystyle{abbrv}

\end{document}